\documentclass[10pt]{IEEEtran}
\usepackage{graphicx}
\usepackage{titlesec}
\usepackage{eso-pic}
\usepackage{float}
\usepackage{algpseudocode}
\usepackage{booktabs}
\usepackage{multirow}
\usepackage{amsmath}
\usepackage{xcolor}
\usepackage{caption}
\usepackage{subcaption}
\usepackage{stfloats}
\usepackage[ruled,vlined,linesnumbered]{algorithm2e}
\usepackage[strings]{underscore}
\usepackage{changepage}
\usepackage{lipsum}
\usepackage{amssymb}
\usepackage{threeparttable}
\usepackage{pdfpages}     
\usepackage{lastpage}     
\usepackage[colorlinks=true, linkcolor=black, citecolor=blue, urlcolor=blue]{hyperref}

\titlespacing*{\subsubsection}
  {0pt}{1.5ex plus 1ex minus .2ex}{1ex plus .2ex}
\titlespacing*{\subsection}
  {0pt}{1.5ex plus 1ex minus .2ex}{1ex plus .2ex}
\titlespacing*{\section}
  {0pt}{1.5ex plus 1ex minus .2ex}{1ex plus .2ex}

\renewcommand\thesection{\arabic{section}}
\renewcommand\thesubsection{\thesection.\arabic{subsection}}

\titleformat{\section}
  [block]
  {\bfseries}
  {\thesection.}
  {1em}
  {}

\titleformat{\subsection}
  {\bfseries}
  {\thesubsection}
  {1em}
  {}

\makeatletter
\renewenvironment{abstract}
{
  \noindent\textbf{Abstract}---\normalfont\normalsize
}{\par\vskip 1em}
\makeatother

\begin{document}

\title{Resilience Evaluation of Kubernetes in Cloud-Edge\\ Environments via Failure Injection}
\author{
  Zihao Chen\textsuperscript{1},
  Mohammad Goudarzi\textsuperscript{1},
  Adel Nadjaran Toosi\textsuperscript{2} \\
  \textsuperscript{1}The Faculty of Information Technology, Monash University, Australia \\
  \textsuperscript{2}School of Computing and Information Systems, The University of Melbourne, Australia
}
\IEEEtitleabstractindextext{%
    \vspace{1em}
    \begin{center}
    \begin{minipage}{0.45\linewidth}
      \centering
      \textbf{Zihao Chen}
    \end{minipage}
    \hfill
    \begin{minipage}{0.45\linewidth}
      \raggedleft
      \textbf{Supervisors:}\\
      Dr. Mohammad Goudarzi\\
      Associate Prof. Adel N. Toosi
    \end{minipage}
    \end{center}
    \vspace{1em}
}
\maketitle

\begin{abstract}
Kubernetes has emerged as an essential platform for deploying containerised applications across cloud and edge infrastructures. As Kubernetes gains increasing adoption for mission-critical microservices, evaluating system resilience under realistic fault conditions becomes crucial. However, systematic resilience assessments of Kubernetes in hybrid cloud-edge environments are currently limited in research. To address this gap, a novel resilience evaluation framework is developed that integrates fault injection tools with automated workload generation for cloud-edge Kubernetes testing. The framework combines multiple fault injection platforms, including Chaos Mesh, Gremlin, and ChaosBlade, with realistic traffic simulation tools to enable automated orchestration of complex failure scenarios. Through this framework, extensive experiments are conducted that systematically target node-level, pod-level, and network failures across cloud and cloud-edge environments. The first comprehensive resilience dataset for hybrid cloud-edge Kubernetes deployments is created, comprising over 30~GB of performance data from 11,965 fault injection scenarios including response times, failure rates, and error patterns. Analysis reveals that cloud-edge deployments demonstrate 80\% superior response stability under network delay and partition conditions, while cloud deployments exhibit 47\% better resilience under bandwidth limitations, providing quantitative guidance for architectural decision-making in cloud-edge deployments.
\end{abstract}

\IEEEpeerreviewmaketitle

\section{Introduction}
\IEEEPARstart{M}{odern} software services are increasingly deployed using containers managed by Kubernetes\footnote{Kubernetes official website: \url{https://kubernetes.io/}}, a widely adopted container orchestration platform~\cite{9826099}. Kubernetes automates the deployment, scaling, and management of containerised workloads, and has experienced rapid adoption in both centralised cloud environments and emerging edge scenarios~\cite{nguyen2022load}. Recent surveys indicate that over 96\% of enterprises already use or plan to adopt Kubernetes in production, highlighting its critical role in modern computing infrastructures~\cite{CNCF2021survey}. Common deployments involve mission-critical applications requiring stringent availability and latency guarantees, such as industrial IoT systems, telecommunications, and healthcare~\cite{Zhang2023}. To satisfy these demands, hybrid cloud-edge architectures have become prevalent, placing application services across centralized clouds and geographically dispersed edge nodes~\cite{cloud-edge}. However, this distributed model introduces increased complexity and vulnerability to faults due to unstable network connectivity and limited edge resources.

Deploying microservices on Kubernetes in cloud-edge environments further complicates resilience management~\cite{10710237}. Unlike traditional monolithic applications, microservices divide functionality into loosely-coupled, independently deployable units, enhancing modularity and scalability but introducing intricate dependencies. Failures in individual microservices may propagate rapidly across dependent services, causing cascading outages~\cite{10137078}. Such cascading effects are especially critical when edge nodes experience network partitions or outages, isolating components and disrupting data flows despite cloud resources remaining unaffected. Kubernetes inherently provides several resilience mechanisms, including automatic pod restarts, workload migration, and health checking of nodes and pods, supported by state management through the distributed key-value store, etcd~\cite{Mutiny}. Nonetheless, these measures have limitations. Studies have demonstrated that minor faults, such as single-bit corruption in etcd or subtle configuration errors, can escalate into significant cluster-wide outages~\cite{Mutiny}. This indicates that Kubernetes-managed systems in distributed cloud-edge scenarios require rigorous, realistic failure assessments beyond conventional reliability tests.

Despite Kubernetes' prominence, systematic studies assessing the resilience of Kubernetes clusters under realistic fault conditions remain sparse~\cite{Mutiny}. Existing research emphasizes component-level functional testing or standard performance metrics without extensively examining application-level resilience metrics like request latency and error rates during faults. Consequently, there is limited empirical evidence contrasting the resilience of monolithic versus microservice architectures in Kubernetes-managed cloud-edge environments~\cite{10037281,hossain2023microservice}. Therefore, understanding the resilience of different application designs under realistic cloud-edge failure scenarios is still an open challenge. 

To address these gaps, we design and evaluate a systematic fault injection framework that assesses Kubernetes resilience at the application level in a cloud-edge context. Leveraging chaos engineering principles, we inject controlled faults into a live Kubernetes cluster and measure their impact on application performance. Our experimental framework combines Chaos Mesh\footnote{Chaos Mesh official website: \url{https://chaos-mesh.org/}}, a native Kubernetes fault injection tool, with Locust\footnote{Locust official website: \url{https://locust.io/}}, a distributed load generator, enabling realistic fault scenarios and concurrent workload simulation. 

The contributions of this paper are summarized as follows:
\begin{itemize}
\item We developed an extensible orchestration framework that automates fault injection, workload generation, and result collection for cloud-edge Kubernetes.  This framework utilizes agentless remote control and modular interfaces for seamless integration with various fault injection and request generation tools. 
\item We performed a comprehensive fault injection study covering a wide array of pod, node, and network failures in both cloud and cloud-edge environments to systematically assess Kubernetes resilience. 
\item We curated a large-scale dataset of resilience metrics from structured fault injection experiments in cloud and cloud-edge settings.  This dataset encompasses response times, success rates, error types, recovery times, and performance degradation patterns under diverse failures like pod crashes, node outages, and network issues. 
\item We conducted a direct empirical comparison of monolithic and microservices architectures using identical fault scenarios within cloud-edge Kubernetes environments. This evaluation reveals their distinct resilience characteristics and highlights trade-offs in stability, recovery, and fault propagation. 
\end{itemize}

The rest of this paper is organized as follows. Section~\ref{background} discusses background information on cloud-edge computing, microservices architecture, and fault injection tools. The proposed resilience evaluation framework is designed in Section~\ref{Proposed framework}. Section~\ref{Experiments and Results} details the experimental design, dataset generation, and results analysis from systematic fault injection in cloud and cloud-edge Kubernetes environments. Finally, Section~\ref{Conclusion} concludes the paper and outlines future research directions.

\section{Background}
\label{background}
This section outlines key concepts and technologies relevant to this work, including cloud-edge computing models, Kubernetes-based microservices orchestration, and existing fault injection approaches for resilience evaluation.
\subsection{Edge-Cloud Computing}
Edge computing extends cloud capabilities by processing data closer to its generation point, meeting stringent requirements for low latency and reduced bandwidth usage~\cite{Satyanarayanan, 9083958}. In practice, cloud and edge resources form a continuum that integrates cloud's scalable computing power with edge nodes' proximity to data sources~\cite{7488250}. This hybrid approach is especially valuable for emerging Internet of Things (IoT) and real-time applications that demand rapid local processing and cannot rely solely on distant cloud data centers~\cite{Zhu2023FLight}. By combining the strengths of both infrastructures, such architectures enhance application responsiveness and network resource efficiency, although at the cost of increased deployment and management complexity~\cite{10.1145/3589639}.

\subsection{Microservices and Container Orchestration}
Kubernetes has emerged as the De facto standard platform for orchestrating containerized applications across clusters of machines~\cite{Chen}. It automates the deployment, scaling, and management of containers (grouped as pods) on distributed worker nodes~\cite{Carrion}. Key features include service discovery, load balancing, and self-healing through mechanisms like health checks and pod restarts~\cite{10294003}. Kubernetes has become pervasive in both cloud and cloud-edge deployments, largely due to its ability to abstract the complexities of managing microservices at scale~\cite{wen2023k8ssim}. By continuously monitoring application state, Kubernetes can automatically react to certain failures, such as rescheduling pods when nodes fail, greatly facilitating the operation of complex systems~\cite{goudarzi2021fogbus2}. However, while Kubernetes provides robust infrastructure-level orchestration, it cannot address application-level resilience challenges arising from microservice interactions' inherent complexity~\cite{wang2024tfddrl}.

Microservice architecture decomposes applications into small, independently deployable services, each handling a specific business function~\cite{10137078}. These services communicate via lightweight APIs to deliver overall functionality, offering benefits in scalability and agility as each microservice can be developed, updated, and scaled individually~\cite{10764926}. However, this approach introduces complexity, as a cloud-edge application may comprise dozens of interdependent microservices deployed across geographically dispersed nodes. This distribution means network calls and partial failures are inherent: a single user request may traverse many services, and failure of one component can impact the whole system if not properly isolated. Despite orchestration advantages, microservice architectures create resilience challenges, as failures within individual services can cascade across dependent components~\cite{10137078}. Empirical studies have revealed significant shortcomings in fault-handling mechanisms, with postmortem analyses frequently uncovering insufficient resilience logic~\cite{7536505}.

In microservices-based edge-cloud systems, failure is inevitable due to the number of components and the unpredictability of the distributed environment~\cite{goudarzi2023muddrl}. Individual services might crash, encounter exceptions, or degrade in performance, while network links between cloud and edge can experience latency or outages~\cite {10037281}. Unlike monolithic systems, where failures affect entire applications, microservices face partial failures that can cascade system-wide if not appropriately handled~\cite{edgeresource}. Cloud-edge deployments amplify this concern as edge nodes may be intermittently connected or resource-constrained, making faults more common~\cite{wang2024reinfog}.
While Kubernetes provides built-in resilience features through liveness/readiness probes and container restarts, studies show these mechanisms miss certain failure modes. Flora et al. examined microservice failures and found that software aging and performance degradation faults often escape detection by Kubernetes probes, as memory leaks may gradually consume resources without triggering immediate crashes~\cite{9996355}. Real-world incidents demonstrate that conventional testing methodologies prove inadequate for identifying complex failure scenarios in distributed microservices, making rigorous resilience testing essential for system validation.

\subsection{Related Work} 
Chaos engineering has emerged as a proactive approach to building resilient systems by deliberately introducing controlled failures~\cite{Mutiny}. Through experiments conducted under ``turbulent'' conditions, engineers uncover unexpected failure scenarios and verify recovery mechanisms' effectiveness. This methodology has gained widespread acceptance, particularly for complex distributed services that undergo frequent changes. It proves especially valuable in hybrid environments that span from centralized cloud infrastructure to resource-constrained edge nodes~\cite{chaosprinciples}.

\subsubsection{Kubernetes Failure Injection}
Kubernetes coordinates multiple essential components (API server, controllers, \textit{etcd}, kubelets) that collectively maintain system health~\cite{edgeresource}. While generally robust against simple failures, the platform contains potential bottlenecks—most notably etcd, the distributed key-value store that maintains all cluster state. Research indicates that because control-plane components operate largely statelessly while etcd centrally stores global state, corruption in etcd data can trigger widespread cluster failures.
Recent work by Barletta et al.\cite{Mutiny} demonstrates that targeted fault injection into Kubernetes' data storage layer (etcd) can reproduce real-world failure patterns, where even single bit-flips may cascade into cluster-wide failures. Their Mutiny framework pioneered control-plane fault injection, revealing that traditional chaos engineering approaches focusing on pod-level disruptions miss critical vulnerabilities in Kubernetes' core infrastructure. However, existing approaches like model-based failure testing~\cite{shen2019model} primarily target application-level services while neglecting systematic evaluation of Kubernetes internal components.
Ergenc et al. \cite{10.1561/1300000074} emphasize the growing need for resilience assessment in edge-cloud applications, noting that current research predominantly focuses on cloud-centric environments while overlooking the complexities of edge coordination. This limitation becomes particularly problematic when considering that van Hoorn et al.~\cite{8539184} found that fault injection and recovery mechanisms for Kubernetes workloads require fundamentally different approaches across distributed deployment scenarios. Basic chaos experiments such as killing pods or introducing network delays typically trigger Kubernetes' self-healing mechanisms effectively. However, more systematic testing approaches are necessary to identify deeper vulnerabilities. While Mutiny enhanced testing by directly injecting state inconsistencies into Kubernetes, revealing subtle failure modes that standard tests often miss~\cite{Mutiny}, it still requires significant expertise to deploy and operate effectively. This limitation also applies to most control-plane testing tools.

\subsubsection{Microservice Failure Injection}
At the application level, fault injection targets microservices and their communication patterns. Building on Netflix's pioneering Chaosmonkey\footnote{Chaosmonkey github: \url{https://netflix.github.io/chaosmonkey/}} approach~\cite{netflixchaos}, modern research has evolved toward sophisticated, targeted approaches addressing random fault injection limitations. Meiklejohn et al.~\cite{10.1145/3472883.3487005} introduced Service-Level Fault Injection Testing (SFIT) through their Filibuster framework, combining static analysis with test generation to systematically explore failure paths between microservices. However, their approach assumes static service interfaces and lacks backend resource failure coverage, limiting applicability to complex microservice dependencies.
Assad et al.~\cite{10.1145/3639478.3640021} extended Filibuster to support fault simulation across SQL and NoSQL databases, enabling systematic resilience verification at the data persistence layer, yet still requiring manual experimental procedure definitions that limit CI/CD scalability. Chen et al.~\cite{10.1109/TDSC.2024.3363902} addressed request-level granularity through their MicroFI framework, providing non-intrusive, prioritized fault injection. Their work highlights that existing tools focusing on service-to-service communication neglect nuanced failure modes from request-level interactions and cascading effects. Yang et al.~\cite{10.1145/3650212.3652131} proposed MicroRes, combining fault injection with performance metrics analysis to quantify resilience through degradation dissemination indexing, but it remains limited to containerized cloud environments without edge-cloud coordination support.
Silva et al.~\cite{10.1145/3555228.3555245} explored distributed fault injection for microservices, revealing that current methods struggle with cross-environment coordination and lack comprehensive observability across distributed failure scenarios, particularly acute in hybrid cloud-edge deployments where failure propagation patterns differ significantly from cloud-only architectures. Network-layer tools like Gremlin\footnote{Gremlin official website: \url{https://www.gremlin.com/}} simulate diverse failure scenarios by intercepting traffic without code modifications, manipulating messages, delays, or API call errors to verify resilience patterns such as retries, fallbacks, and circuit breakers. However, these approaches primarily focus on isolated environment testing and lack integrated workload generation capabilities necessary for comprehensive resilience evaluation under realistic operational conditions in cloud-edge environments.
\subsubsection{Fault Injection Tools and Frameworks}
The Kubernetes fault injection landscape has evolved significantly, with tools exhibiting varying capabilities across critical dimensions as illustrated in Table~\ref{tab:framework-comparison}. 

\begin{table}[h]
\scriptsize
\begin{minipage}{0.85\linewidth}
\centering
\caption{Comparison of Kubernetes Fault Injection Tools and Frameworks}
\label{tab:framework-comparison}
\renewcommand{\arraystretch}{1.1}
\setlength{\tabcolsep}{2pt}
\resizebox{1.15\linewidth}{!}{
\begin{tabular}{|l|c|c|c|c|c|c|}
\hline
\textbf{Tool/Framework} & \textbf{Pod} & \textbf{Net} & \textbf{Ctrl} & \textbf{Load} & \textbf{App} & \textbf{Cloud} \\
\textbf{} & \textbf{Faults} & \textbf{Faults} & \textbf{Plane} & \textbf{Gen.} & \textbf{Metrics} & \textbf{Edge} \\
\hline
Chaosmesh~\cite{chaosmesh}       & \checkmark & \checkmark & $\times$ & $\times$ & $\times$ & \checkmark \\
Litmuschaos~\cite{litmuschaos}     & \checkmark & \checkmark & $\times$ & $\times$ & $\times$ & \checkmark \\
Gremlin~\cite{7536505}         & \checkmark & \checkmark & $\times$ & $\times$ & \checkmark & $\times$ \\
Chaosblade~\cite{chaosblade2019}       & \checkmark & \checkmark & $\times$ & $\times$ & $\times$ & \checkmark \\
Chaostoolkit~\cite{chaostoolkit2016}   & \checkmark & \checkmark & $\times$ & $\times$ & $\times$ & \checkmark \\
Powerfulseal~\cite{powerfulseal2020}     & \checkmark & \checkmark & $\times$ & $\times$ & $\times$ & \checkmark \\
chaoskube~\cite{chaoskube2016}       & \checkmark & $\times$   & $\times$ & $\times$ & $\times$ & \checkmark \\
Mutiny~\cite{Mutiny}         & \checkmark & \checkmark & \checkmark & $\times$ & $\times$ & $\times$ \\
\textbf{Our Work} & \textbf{\checkmark} & \textbf{\checkmark} & \textbf{\checkmark} & \textbf{\checkmark} & \textbf{\checkmark} & \textbf{\checkmark} \\
\hline
\end{tabular}
}
\begin{adjustwidth}{1.6em}{2pt}
\vspace{1em}
\scriptsize

\textsc{Pod Faults}: Pod/Node level fault injection. \\
\textsc{Net Faults}: Network fault injection capability. \\
\textsc{Ctrl Plane}: Control-plane fault injection. \\
\textsc{Load Gen.}: Workload generation integration. \\
\textsc{App Metrics}: Application performance metrics. \\
\textsc{Cloud-Edge}: Cloud-edge coordination support.

\end{adjustwidth}
\end{minipage}
\end{table}

Open-source platforms like Chaos Mesh and LitmusChaos\footnote{LitmusChaos official website: \url{https://litmuschaos.io/}} establish infrastructure-level testing foundations, leveraging Kubernetes-native Custom Resource Definitions (CRDs) to orchestrate pod failures and network disruptions, yet primarily target layers where Kubernetes provides robust self-healing, leaving control-plane vulnerabilities unexplored. Bagehorn et al.~\cite{10.1145/3551349.3559503} developed an automated fault injection platform for Artificial Intelligence for IT Operations (AIOps), model training, demonstrating experiment management simplification potential but remaining focused on single-environment deployments without multi-environment orchestration capabilities. The ORCAS framework~\cite{8539184} leverages architectural knowledge to automatically generate fault injection experiments, improving testing efficiency over random approaches, but struggles with dynamic cloud-edge deployments where service topology and resource constraints vary significantly. Norris et al.~\cite{10.1145/3532194} explored multilevel fault injection in IoT-edge systems and revealed critical limitations in existing frameworks. Their study highlights that most cloud-centric tools fail to adequately address edge-specific challenges, such as resource constraints, intermittent connectivity, and heterogeneous hardware platforms. This underscores the need for edge-aware fault injection approaches in distributed edge-cloud environments.

Recent comprehensive surveys~\cite{9688292,owotogbe2025chaos} identify that while chaos engineering practices have matured, most frameworks lack automated integration capabilities and cross-layer testing support. Sile et al.~\cite{sile2023chaos} noted that chaos orchestration for cloud-native applications remains fragmented, with tools targeting specific layers without unified stack management. The ecosystem shows distinct specialization: ChaosBlade\footnote{Chaosblade official website: \url{https://chaosblade.io/en/}} excels with fine-grained kernel-level fault injection across 200+ failure scenarios, Chaos Toolkit\footnote{Chaostoolkit official website: \url{https://chaostoolkit.org/}} offers platform-agnostic testing through structured YAML definitions, while lightweight tools like PowerfulSeal\footnote{Powerfulseal official website: \url{https://powerfulseal.github.io/powerfulseal/}}
and chaoskube\footnote{Chaoskube official website: \url{https://github.com/linki/chaoskube}} focus on targeted pod disruptions, trading complexity for ease of use. Gremlin distinguishes itself with enhanced observability features, yet Higgins et al.~\cite{10.1145/3625549.3658827} note that even advanced tools struggle with automated chaos experimentation at scale.

Joshua et al. \cite{owotogbe2024multivocal} emphasize that scalable chaos testing infrastructure remains challenging, particularly for frameworks supporting both centralized cloud resources and distributed edge nodes. Borges et al.~\cite{borges2025observability} highlight that observability integration, which is critical for understanding failure impact, remains insufficient in most tools, limiting production effectiveness. A critical limitation pervades mainstream tools: their inability to target control-plane components represents a significant resilience evaluation blind spot. Moreover, most solutions lack streamlined deployment and one-click execution capabilities, requiring complex setup procedures and manual fault scenario orchestration, creating adoption barriers and CI/CD integration complications. Mutiny~\cite{Mutiny} addressed control-plane testing gaps through etcd corruption and API server disruption testing, but still requires significant deployment and operational expertise.

Our framework represents the next evolutionary step in Kubernetes resilience testing, uniquely combining control-plane fault injection with integrated workload generation and comprehensive metrics collection. By integrating Chaos Mesh for fault injection with Locust for realistic traffic simulation, we enable precise measurement of application degradation during fault conditions, which is critical for cloud-edge environments where reliability requirements are heightened by distributed architectures~\cite{chaosmesh}. Unlike existing solutions, our framework provides a streamlined, one-click deployment experience that simplifies resilience testing in production-like environments, making comprehensive fault injection accessible to development teams without specialized chaos engineering expertise. This integration delivers the end-to-end resilience evaluation missing in existing solutions, with structured orchestration and multi-level observability spanning both cloud and edge components. The complete implementation is publicly available on GitHub\footnote{\url{https://github.com/dylanC777/cloud-edge-k8s-resilience}} for academic and research use.

\subsection{Research Gap and Contribution}
Despite advances in fault injection for cloud-native systems, holistic resilience testing across cloud-edge environments remains lacking. Existing studies primarily focus on isolated cloud environments without examining failure propagation across the cloud-edge continuum. Additionally, no prior work has systematically compared monolithic and microservice architectures under identical fault conditions in such hybrid deployments.

Our research addresses these gaps through systematic fault injection experiments in a Kubernetes-managed cloud-edge testbed. We conduct chaos experiments spanning both environments to observe system behavior under various failure scenarios, including pod, node, and network failures. By deploying monolithic and microservice versions of an application under identical conditions, we provide the first quantitative comparison of their resilience characteristics.

Key contributions include: (1) a novel testing framework that automates fault injection and workload generation, (2) systematic experiments comparing architectural resilience under identical fault conditions, and (3) a detailed dataset of resilience metrics under varied fault conditions. No previous research has provided such evaluation matrices for cloud-edge Kubernetes deployments. Our approach yields valuable insights into architectural resilience in hybrid environments, informing more robust cloud-edge application design and improved fault-tolerance strategies. This work establishes the foundation for evidence-based deployment decisions in distributed cloud-edge computing environments. The findings provide practitioners with quantitative guidance for selecting optimal deployment strategies based on specific fault tolerance requirements and operational constraints.

\section{Proposed Framework}
\label{Proposed framework}

\begin{figure*}[t]
    \centering
    \includegraphics[width=\textwidth]{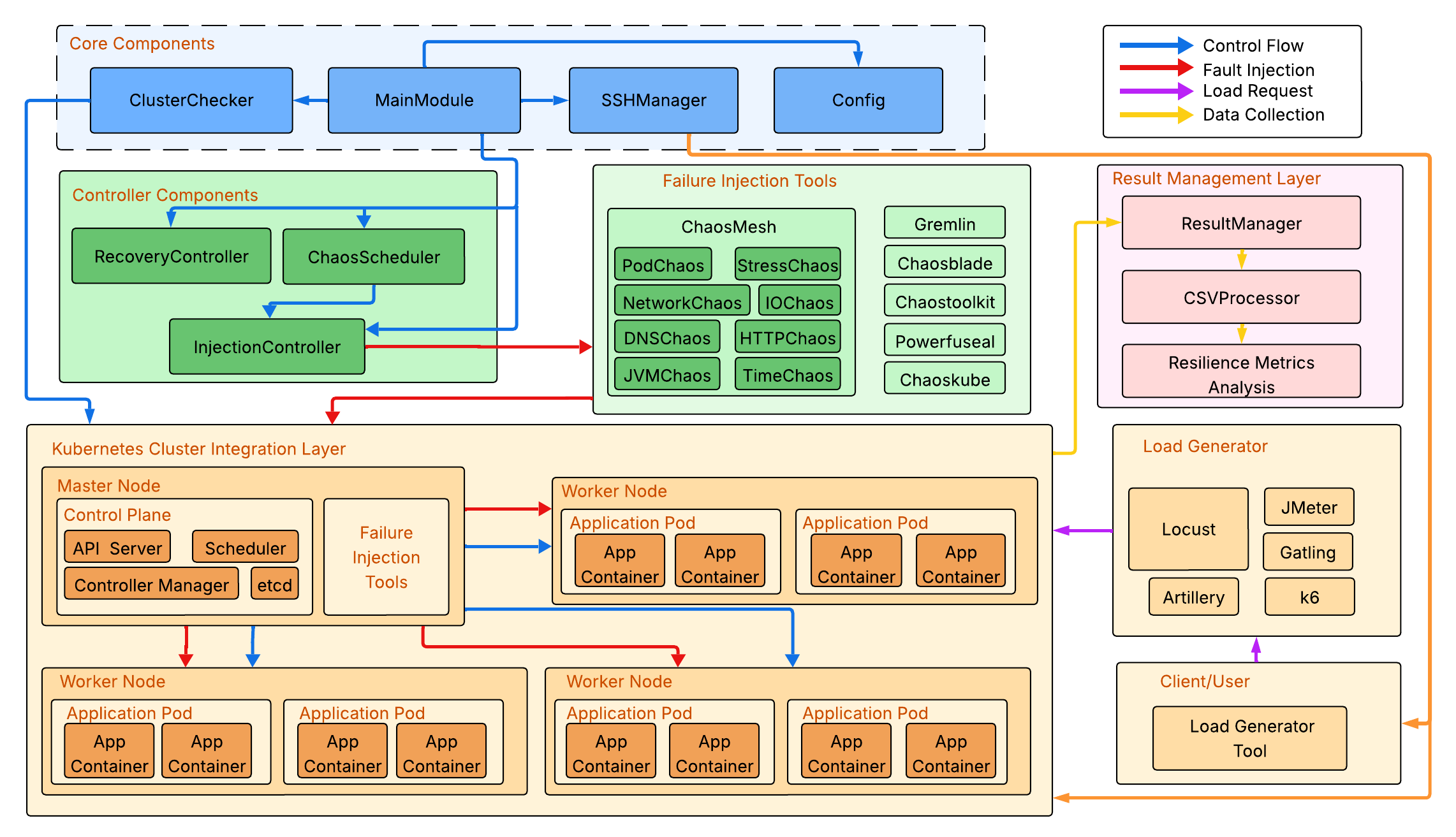}
    \caption{Layered architecture of the proposed resilience evaluation framework, including the orchestrator, experiment controllers, external tools, and the target Kubernetes cluster. Blue arrows indicate control flow, red for fault injection, purple for load requests, and orange for data collection.}
    \label{fig:framework-architecture}
\end{figure*}

To systematically evaluate Kubernetes resilience in cloud-edge environments, we design a unified framework grounded in chaos engineering principles. Chaos engineering has proven to be a highly effective strategy for enhancing system reliability by introducing controlled failures and conducting deliberate stress experiments~\cite{10903891}. 

Our framework provides a centralized orchestration mechanism that manages fault injection studies across both cloud and edge nodes, enabling systematic evaluation of application resilience in heterogeneous deployments. By unifying the management of diverse tools and distributed infrastructure under a single control plane, our framework significantly simplifies the complexity traditionally associated with cloud-edge chaos experiments. The design emphasizes automation, reproducibility, and seamless integration with existing chaos engineering tools, while supporting comparative evaluation of different application architectures including monolithic and microservices-based systems.

To realize these objectives, the framework adopts a modular, layered architecture that separates key functions and supports flexible experimentation with failure scenarios. It comprises six layers, illustrated in Figure~\ref{fig:framework-architecture}, each with distinct responsibilities and designed for seamless integration through clear interfaces. This modularity allows new tools or fault types to be added without changing the core orchestration logic, ensuring adaptability to future chaos engineering practices.

\subsection{Framework Architecture}
\label{3.1}
The following subsections detail the specific implementation and functionality of each component layer, examining their individual responsibilities, interaction mechanisms, and contribution to the overall experimental workflow. We present the core infrastructure components that establish the foundation for distributed chaos engineering, followed by the specialized control and management layers that orchestrate complex experimental scenarios.

\subsubsection{Core Components} 
The core components form the foundation of the framework, providing essential services for configuration management, connectivity, health monitoring, and orchestration. These components work together to establish the fundamental infrastructure required for systematic chaos engineering experiments across distributed cloud-edge environments.

The \textit{Cluster Checker} component serves as the health monitoring subsystem, ensuring the Kubernetes cluster maintains operational stability before, during, and after chaos experiments. This component implements comprehensive health validation mechanisms that monitor all Kubernetes nodes to ensure they remain in the ready state through direct API queries. The system identifies active Chaos Mesh schedules to prevent conflicting experiments, while validating that all application pods are running and ready within specified namespaces using readiness fraction parsing. This component supports automated recovery waiting with configurable retry intervals, and provides clean state restoration capabilities through parallel deployment restart operations. The health check process follows a systematic approach where node readiness, chaos schedule absence, and pod health are validated collectively, ensuring experiments only proceed when the cluster is in a stable state.

The \textit{Main Module} is the central orchestration engine, coordinating distributed framework components throughout the experimental lifecycle while integrating secure remote connectivity and centralised parameter management through a unified interface. This orchestrator employs advanced configuration parsing to extract experimental parameters and infrastructure specifications from hierarchical YAML definitions, subsequently establishing encrypted SSH communication channels for agentless operation across cloud-edge topologies. The module implements adaptive experimental sequencing that manages synchronised chaos injection, concurrent load generation, and result aggregation while supporting dynamic configuration of execution threads and timeout thresholds through comprehensive parameter orchestration and template-driven experiment design. 

The system incorporates intelligent error recovery mechanisms with configurable retry policies spanning request transmission failures, fault injection errors, load generation failures, and cluster validation operations, alongside systematic inter-experiment recovery protocols encompassing deployment restoration, health verification, and resource cleanup to ensure experimental isolation. Experimental execution follows a structured pipeline where each iteration applies parameterised chaos configurations with dynamic parameter substitution, executes multi-threaded load testing with adaptive retry logic, and aggregates performance metrics to facilitate reproducible experimental campaigns across diverse deployment scenarios.

\subsubsection{Controller Components and Failure Injection System}
The \textit{controller layer} bridges the orchestration engine with the underlying fault injection mechanisms, providing specialized management functions that ensure systematic and safe execution of chaos experiments. This layer encompasses both recovery management capabilities and integrated fault injection systems that work cohesively to maintain experimental integrity while delivering comprehensive failure simulation across distributed cloud-edge environments.

The \textit{Recovery Controller} implements systematic post-experiment cleanup and state restoration procedures that ensure experimental isolation and baseline consistency between test iterations. This component manages configurable stabilisation periods between experiments, automatically coordinating deployment restart operations to achieve clean state initialisation for subsequent experimental runs. The controller incorporates verification mechanisms that confirm successful system restoration before permitting progression to subsequent experiments, thereby preventing cascading failures and maintaining experimental validity across extended test campaigns. These safety mechanisms align with chaos engineering best practices, which emphasise the importance of controlled experimentation~\cite{11112}.

The \textit{Chaos Scheduler} coordinates temporal aspects of fault injection across diverse fault types and experimental scenarios. This component supports both isolated single-fault experiments and coordinated multi-fault scenarios, managing fault intensity progression through systematic percentage-based resource targeting across four distinct levels. The scheduler provides precise temporal control over fault application and removal phases, ensuring consistent experimental conditions and reproducible results throughout complex experimental sequences. 

The \textit{Injection Controller} integrates directly with Chaos Mesh as the primary fault injection platform, providing a unified interface for comprehensive failure simulation capabilities. This integration leverages Kubernetes Custom Resource Definitions to enable native container orchestration system interaction, supporting extensive fault categories including container termination, pod elimination, network delay injection, network loss simulation, and bandwidth throttling operations. This controller implements fine-grained targeting mechanisms with percentage-based resource selection, enabling precise control over fault scope and intensity across distributed system components. While the current implementation focuses primarily on Chaos Mesh integration, the controller architecture maintains extensibility provisions for future integration with additional chaos engineering platforms, including Gremlin, ChaosBlade, and other specialised fault injection tools, ensuring framework adaptability to evolving chaos engineering ecosystems.

\begin{figure*}[tb]
    \centering
    \includegraphics[width=1\textwidth]{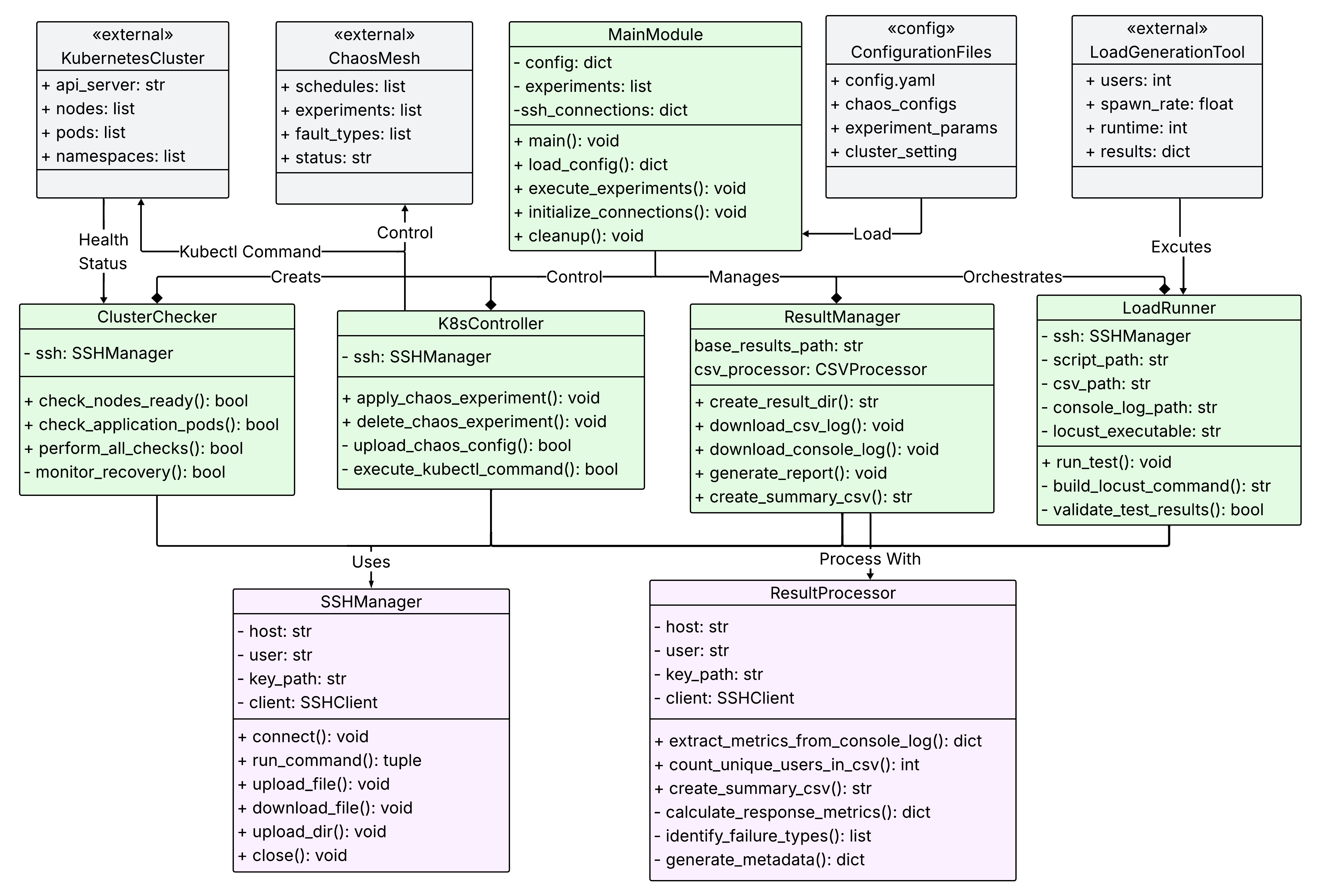}
    \caption{Internal design of core framework components (UML class diagram).}
    \label{fig:classdiagram}
\end{figure*}

\subsubsection{Load Generation System}
The \textit{Load Generation System} provides sophisticated workload simulation capabilities to recreate realistic application usage patterns during chaos experiments, enabling comprehensive performance evaluation under controlled stress conditions. This component supports multiple operational modes to accommodate diverse testing scenarios and application behavioural characteristics across cloud-edge deployments. The framework integrates Locust as the primary load generation platform, leveraging its distributed testing capabilities and Python-based scripting flexibility. Through this integration, the system supports three distinct operational modes:
\begin{itemize}
    \item \textbf{Piggyback mode}: Executes background traffic patterns with periodic bursts to reflect realistic application usage variations.
    \item \textbf{Concurrent mode}: Provides unlimited request rate capabilities for maximum throughput evaluation.
    \item \textbf{Constant rate mode}: Implements steady request patterns with configurable intervals for typical application usage simulation.
\end{itemize}

The Locust integration incorporates advanced configuration capabilities, including timeout management mechanisms, dynamic environment variable injection, and adaptive path configuration based on selected testing modes. The implementation provides comprehensive error detection and retry mechanisms that analyze test execution through status monitoring and log analysis to ensure reliable completion and accurate data collection. The system architecture maintains extensibility provisions for future integration with additional load generation tools, including JMeter, Artillery, and other specialized performance testing platforms, enabling practitioners to select optimal tools based on specific experimental requirements and application characteristics while maintaining consistent experimental orchestration and result collection capabilities.

\subsubsection{Result Manager Layer}
The result management system provides comprehensive data collection and analysis capabilities for distributed cloud-edge experimental environments. This layer implements centralized monitoring mechanisms that continuously assess system health through periodic API queries, providing early warning systems for severe degradation while maintaining observational capabilities under stress conditions. The data collection subsystem implements automated retrieval and aggregation of experimental artifacts across heterogeneous infrastructures. The system retains comprehensive metadata preservation to ensure reproducibility, capturing experimental context, parametric configurations, and temporal sequences. Advanced handling mechanisms employ adaptive sampling strategies for large-scale outputs, optimizing network resource utilization while preserving critical diagnostic information. The analytical processing framework utilizes dedicated metric extraction engines to generate performance summaries and structured reports. This system implements correlation analysis between fault injection timing and observed effects, enabling precise identification of failure propagation patterns. The framework maintains standardized data formats supporting immediate analysis and longitudinal trend evaluation, facilitating systematic comparison between fault scenarios and architectural configurations through consistent metric calculation methodologies.

\subsubsection{Kubernetes Cluster Integration Layer}
The \textit{Kubernetes Cluster Integration Layer} provides abstracted cluster management capabilities that enable seamless framework interaction with distributed Kubernetes environments through standardized interfaces, handling operational complexity while maintaining consistent experimental control across diverse deployment scenarios. 

This layer implements comprehensive chaos experiment management through YAML-based deployment procedures that apply Chaos Mesh configurations via cluster API interactions alongside shell script execution support for custom scenarios requiring extended fault conditions. The system incorporates systematic cleanup operations, ensuring experimental environments return to baseline conditions between test runs through schedule deletion and resource management procedures. The implementation provides flexible deployment support through automatic configuration type detection, determining whether to apply YAML configurations through cluster APIs or execute background shell scripts, thereby accommodating varying experimental requirements while maintaining consistent operational procedures and result collection capabilities across heterogeneous cloud-edge infrastructures.

\begin{figure*}[t]
    \centering
    \includegraphics[width=1\textwidth]{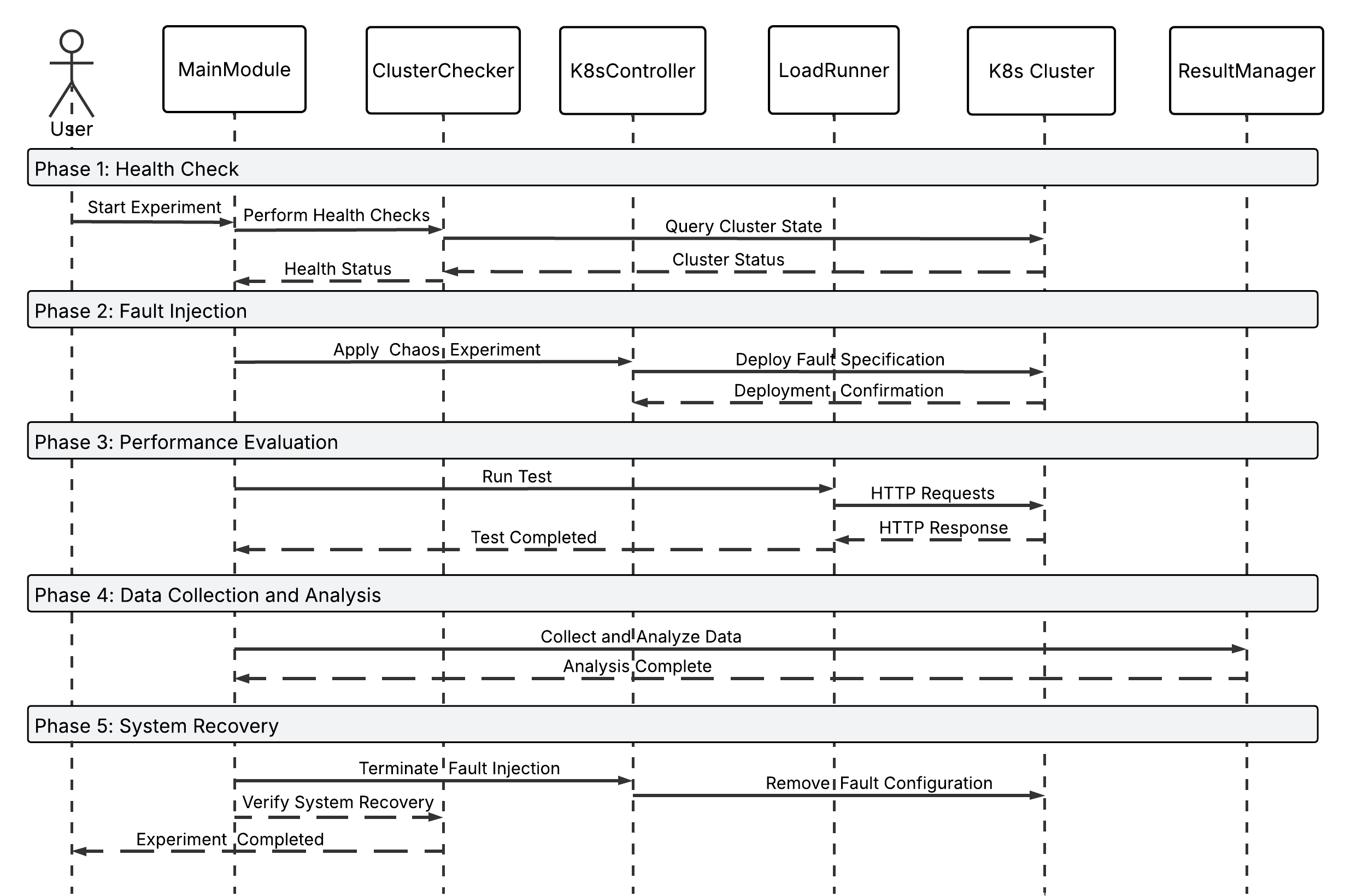}
    \caption{UML sequence diagram of the framework’s five-phase cloud-edge Kubernetes resilience experiment, illustrating the interactions from pre-experiment health checks through fault injection, workload execution, data analysis, and final system recovery.}
    \label{fig:sequencediagram}
\end{figure*}

\subsubsection{Modular Architecture and Design Patterns}

Our framework's static design leverages object-oriented principles that promote modularity and extensibility in heterogeneous cloud-edge settings, as depicted in Figure~\ref{fig:classdiagram}. 

At the centre is \textit{Main Module}, which simultaneously fulfils the roles of \emph{Facade} and \emph{Mediator}, by exposing a single entry point that internalises the canonical experimental workflow configuration loading, baseline health auditing, fault activation, workload generation, metric collection, and post-experiment recovery. The \textit{Main Module} applies the \emph{Template Method} pattern to guarantee a uniform life-cycle across all experiments while shielding higher-level components from coordination complexity. The supporting classes each embody a well-defined concern in accordance with the Single Responsibility Principle.  \textit{Cluster Checker} conducts pre-fault and post-fault validation of node, namespace, and application health.  \textit{K8s Controller} translates declarative fault profiles into Chaos Mesh custom resources or Gremlin scripts and supervises their execution.  \textit{Load Runner} constructs parameterised Locust commands and initiates traffic from both cloud and edge clients.  \textit{Result Manager} retrieves CSV logs, console traces, and auxiliary metadata, then delegates metric extraction to  \textit{Result Processor}.  All components rely on a shared \textit{SSH Manager}, implemented as a \emph{Singleton} that maintains a pool of long-lived, keep-alive connections; this agent-less strategy minimises TCP handshake overhead and provides automatic reconnection in the presence of intermittent edge links. To insulate the framework core from third-party dependencies, external systems, including the Kubernetes API server, Chaos Mesh controller, and alternative load generators, are accessed exclusively through dedicated adapter classes.  Adding a new fault-injection engine or workload driver therefore requires only the implementation of an additional adapter, leaving the remainder of the framework untouched.  This plug-in capability preserves conceptual coherence with the five-layer architecture outlined in Section~\ref{3.3} and ensures that the static design can evolve alongside advances in chaos-engineering tooling without compromising the integrity or maintainability of the overall system.

\subsection{Execution Workflow and Orchestration}
\label{3.3}

Having established the modular architecture and design patterns, we now examine how these components coordinate during actual experiment execution. The end-to-end workflow unfolds in five tightly coordinated phases, shown in the UML sequence diagram of Figure~\ref{fig:sequencediagram}. The experiment is triggered by the \emph{User}. The \textit{Main Module} then initiates a pre-flight validation. During Phase 1, this \textit{Main Module} issues a synchronous call to \textit{Cluster Checker}, which in turn queries the Kubernetes API server to verify node readiness, \emph{namespace} consistency, and pod-level liveness. Only when a positive health status is returned does the orchestrator advance to the fault stage, thereby guaranteeing a stable baseline and eliminating hidden pre-existing anomalies. \emph{Phase 2} then centres on controlled fault activation, where \textit{Main Module} delegates the operation to \textit{K8s Controller}, which applies a \emph{Chaos Mesh} custom resource to the cluster; the controller awaits an asynchronous confirmation event signalling that the fault specification has been successfully deployed. This explicit handshake not only bounds the activation latency to a single round-trip but also permits the orchestrator to record precise fault-on timestamps that later support correlation analysis.

With the perturbation in effect, \emph{Phase 3} launches workload generation. \textit{Main Module} coordinates with \textit{Load Runner}, which constructs a parameterised Locust command according to the current experiment configuration and streams HTTP requests towards the instrumented services. The bi-directional message exchange, comprising normal responses, error codes, and time-out events persists for the predefined runtime window, during which component lifelines remain active to capture transient failures. Because request traffic and injected faults execute concurrently, the framework exposes latency amplification, throughput collapse, and cascading failure phenomena that would otherwise be masked in sequential test designs.

Upon completion of the workload window, \emph{Phase 4} is invoked automatically, where the \textit{Main Module} coordinates with the \textit{Result Manager} to collect and analyze experimental data. The \textit{Result Manager} retrieves raw CSV logs and console traces via the shared SSH channel, extracts salient metrics through the \textit{Result Processor}, and consolidates them into a structured report with performance summaries and resilience indicators. The data pipeline exploits incremental transfer and on-the-fly compression to mitigate bandwidth contention across the cloud–edge link, ensuring that large artefacts can be collected without impeding subsequent experiments. Finally, \emph{Phase 5} orchestrates system recovery, where \textit{K8s Controller} removes the fault specification, after which \textit{Cluster Checker} re-enters its validation loop to confirm that all resources have returned to a Ready state before \textit{Main Module} marks the experiment as complete and proceeds to the next parameter set. This bounded-recovery protocol with configurable back-off and retry semantics prevents residual side effects and preserves experimental orthogonality.

Compared with ad-hoc scripting approaches, this orchestrated, agent-less workflow unifies health validation, fault activation, workload execution, data acquisition, and recovery verification behind a single \emph{façade}. It therefore reduces operational overhead, minimises human error, and delivers repeatable conditions for statistically rigorous resilience assessment across heterogeneous cloud-edge deployments.

\section{Experiments and Result Analysis}
This section presents systematic resilience evaluation experiments across cloud and cloud-edge Kubernetes environments, covering experimental setup, dataset, and result analysis.

\label{Experiments and Results}
\subsection{Experimental Setup and Design}
This subsection presents our resilience evaluation experiments' systematic design and configuration across cloud and cloud-edge Kubernetes environments. The setup involves 11,965 distinct experimental scenarios that systematically vary deployment architectures, fault types, workload patterns, and infrastructure configurations to enable thorough resilience characterization and comparative analysis between different deployment strategies

\label{Experimental}
\subsubsection{Infrastructure Setup}
Our experiments systematically evaluate Kubernetes resilience across four distinct deployment scenarios using two Kubernetes clusters: a 4-node cluster and an 8-node cluster. Each cluster is configured to operate in both pure cloud and cloud-edge hybrid modes, resulting in four environmental configurations that reflect real-world deployment patterns.

The following table~\ref{tab:infrastructure-config} details the complete technical specifications of our experimental infrastructure, including the Kubernetes environment, cloud platform characteristics, and hardware configurations that ensure consistent and reproducible experimental conditions:

\begin{table}[H]
\centering
\caption{Infrastructure Configuration Summary}
\label{tab:infrastructure-config}
\begin{tabular}{ll}
\toprule
\textbf{Component} & \textbf{Specification} \\
\midrule
\textbf{Kubernetes Version} & v1.27.4, default kubeadm configuration \\
\midrule
\multirow{3}{*}{\textbf{Cluster Structure}} & 1 control plane node \\
& Worker nodes \\
& External monitoring node \\
\midrule
\textbf{Network Manager} & Weave Net \\
\midrule
\textbf{Cloud Platform} & NeCTAR Research Cloud \\
\midrule
\textbf{Instance Types} & m3.large (8 vCPUs, 16GB RAM) \\
\midrule
\textbf{Operating System} & NeCTAR Ubuntu 22.04 LTS (Jammy) amd64 \\
\midrule
\textbf{Storage} & 30GB SSD \\
\midrule
\textbf{Container Runtime} & containerd v1.7 \\
\bottomrule
\end{tabular}
\end{table}

To provide a comprehensive view of our deployment configurations and network simulation parameters, the table~\ref{tab:cluster-deployment} summarizes the specific arrangements for both cluster sizes and their corresponding cloud-edge hybrid configurations:
\begin{table}[H]
\centering
\caption{Cluster Deployment Configurations}
\label{tab:cluster-deployment}
\begin{tabular}{lll}
\toprule
\textbf{Cluster Type} & \textbf{Cloud Mode} & \textbf{Cloud-Edge Hybrid Mode} \\
\midrule
\multirow{3}{*}{\textbf{4-Node Cluster}} & 4 worker nodes with & 3 cloud + 1 edge node \\
& standard network & 200\,ms latency (±10\%) \\
& & 10\% network loss \\
\midrule
\multirow{3}{*}{\textbf{8-Node Cluster}} & 8 worker nodes with & 5 cloud + 3 edge nodes \\
& standard network & 200\,ms latency (±10\%) \\
& & 10\% network loss \\
\bottomrule
\end{tabular}
\end{table}

The cloud environment provides stable, high-bandwidth connectivity with minimal latency, representing optimal datacenter conditions. Edge simulation implements realistic edge conditions through controlled network impairments: 200\,ms base latency with 10\% random variation (180--220\,ms range) to simulate network jitter, 10\% packet drop rate to represent unstable edge connectivity, and variable throughput limitations typical of edge deployments. All nodes are virtual machines with consistent hardware specifications, ensuring a reliable baseline and allowing for realistic resource contention under stress. The edge simulation parameters are based on real-world measurements from industrial IoT and remote edge deployments. This dual-cluster, dual-environment approach allows for the isolation of scale and distribution impacts on system resilience, while the network impairments provide realistic testing conditions for cloud-edge scenarios.

\subsection{Application Configuration}
To assess architectural resilience, we deployed two applications representing contrasting design philosophies and subjected both architectures to identical fault scenarios across our experimental infrastructure. This provided empirical evidence on resilience trade-offs, an aspect that remains underexplored in existing literature.
 
1. \textit{Monolithic Application (Image-Detection)}: A deep-learning image classification service representative of latency-sensitive edge workloads common in industrial IoT and video analytics. Its monolithic nature consolidates functionality, simplifying deployment but potentially introducing a single point of failure, as all application components are packaged and deployed within a single Kubernetes pod.

\begin{figure}[H]
    \centering
    \includegraphics[width=\columnwidth]{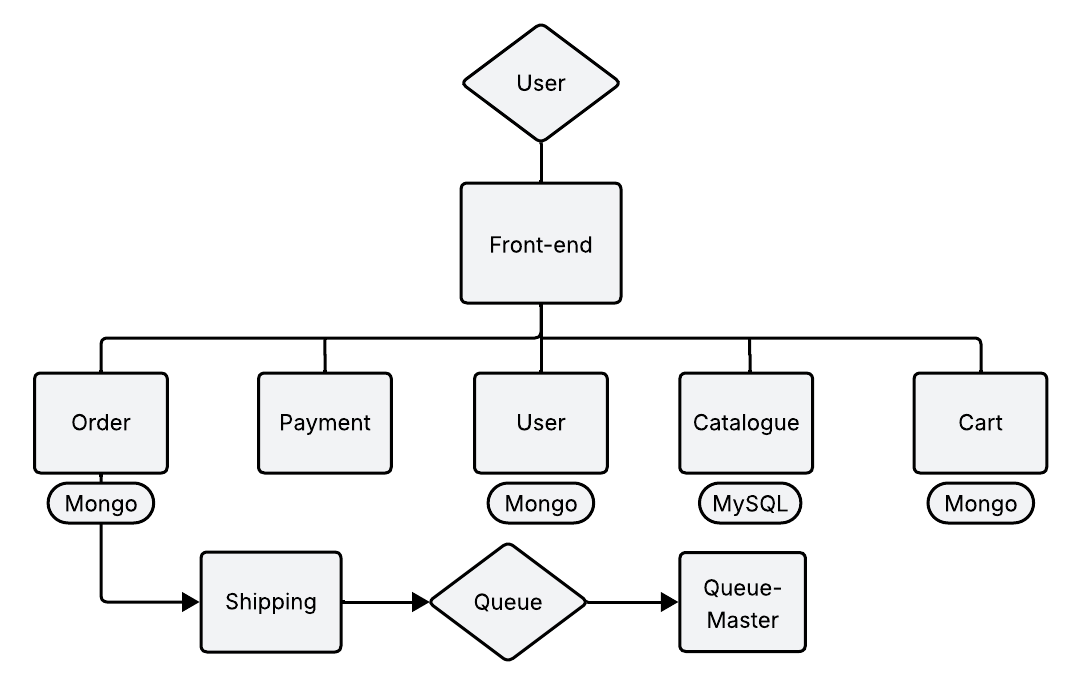}
    \caption{Sock Shop Microservice Benchmark}
    \label{fig:sock-shop}
\end{figure}

2. \textit{Microservices Application (Sock-Shop\footnote{SockShop: \url{https://github.com/ocp-power-demos/sock-shop-demo/tree/main}})}: A microservices-based e-commerce platform composed of 13 interdependent services, including frontend, catalogue, orders, and payment components~\cite{sockshop}. As illustrated in Figure~\ref{fig:sock-shop}, this application typifies modern cloud-native designs with distributed, loosely-coupled services. While offering scalability and fault isolation, it introduces complex inter-service dependencies and risks of cascading failures.

\subsubsection{Experiment Configuration}
Our evaluation explores resilience across multiple dimensions to generate a comprehensive dataset for cloud-edge Kubernetes deployments. Table~\ref{tab:experimental-design} summarises our experimental configuration, which systematically combines workload patterns, fault injection scenarios, and infrastructure variations.

\begin{table}[H]
\centering
\caption{Experimental Design Configuration}
\label{tab:experimental-design}
\begin{tabular}{ll}
\toprule
\textbf{Component} & \textbf{Configuration} \\
\midrule
\multirow{3}{*}{\textbf{Workload Patterns}} & Constant: Steady 5 req/s per thread \\
& Concurrent Burst: Sudden traffic spikes \\
& Piggyback: Background + periodic bursts \\
\midrule
\textbf{Thread Configurations} & 1, 2, 4, 8, 16 threads \\
\textbf{Timeout Range} & 1--10 seconds \\
\midrule
\multirow{6}{*}{\textbf{Fault Types}} & Container termination \\
& Pod termination \\
& Network delay injection \\
& Network loss simulation \\
& Bandwidth throttling \\
& CPU stress testing \\
& Node termination\\
\midrule
\textbf{Fault Intensity} & 25\%, 50\%, 75\%, 100\% of resources \\
\midrule
\textbf{Applications} & Monolithic (Image Detection) \\
& Microservices (Sock Shop) \\
\midrule
\textbf{Infrastructure} & 4-node cluster \\
& 8-node cluster \\
\midrule
\textbf{Deployment Modes} & Pure cloud environment \\
& Cloud-edge hybrid environment \\
\bottomrule
\end{tabular}
\end{table}

We employ Locust to generate realistic workload patterns that represent typical production scenarios. The systematic variation of client-side timeouts (1-10 seconds) enables identification of optimal configurations for different architectural patterns, as monolithic applications typically favour shorter timeouts while microservices benefit from longer timeouts to accommodate inter-service communication latency.

Our fault injection leverages Chaos Mesh to implement controlled failures representing common production incidents. Each fault type is executed at four intensity levels, with the proportional approach ensuring generalizability across deployment scales. For example, network delay faults are mapped to specific latencies (25\%=100ms, 100\%=1000ms), while bandwidth throttling applies corresponding limits (25\%=10Mbps, 100\%=1Mbps). 
Table~\ref{tab:fault-example} illustrates a representative fault injection configuration targeting 75\% of microservice containers.

\begin{table}[h]
\centering
\caption{Fault Injection Configuration Example}
\label{tab:fault-example}
\begin{tabular}{ll}
\toprule
\textbf{Parameter} & \textbf{Value} \\
\midrule
Action & Container-kill \\
Mode & fixed-percent \\
Value & 75 \\
Targets & \{carts, catalogue, user, Payment, Shipping\} \\
Duration & 3s \\
Trigger Frequency & every 3s \\
\bottomrule
\end{tabular}
\end{table}

The systematic variation of workload patterns, timeout settings, and fault intensities ensures comprehensive coverage of realistic operational conditions across both architectural types and deployment scenarios.

\subsection{Dataset}

\begin{table*}[t]
\centering
\caption{Dataset Summary}
\label{tab:cluster-arch-summary}
\small
\setlength{\tabcolsep}{4pt}
\begin{tabular}{lcccc}
\toprule
\textbf{Field} & \textbf{Cluster 4·Image-Detection} & \textbf{Cluster 4·Sock-Shop} & \textbf{Cluster 8·Image-Detection} & \textbf{Cluster 8·Sock-Shop} \\
\midrule
Experiment count           & 3\,332 & 3\,000 & 2\,792 & 2\,840 \\
Total requests (million)   & 16.0   & 14.0   & 12.0   & 15.0   \\
Approx.\ raw log size (GB) & 8.0    & 7.0    & 6.0    & 8.0    \\
Mean Response Time (ms)$^\dagger$ & 873    & 1\,470  & 831    & 869    \\
P95 Response Time (ms)$^\ddagger$ & 4\,156 & 4\,948  & 3\,482  & 4\,516  \\
Failure rate (\%)$^\S$ & 18.8   & 43.4   & 13.1   & 23.1   \\
\bottomrule
\multicolumn{5}{l}{\scriptsize $^\dagger$Mean response time = average per-request response time across all experiments.}\\
\multicolumn{5}{l}{\scriptsize $^\ddagger$P95 response time = 95th percentile of per-request response times across all experiments.}\\
\multicolumn{5}{l}{\scriptsize $^\S$Failure rate = number of failed experiments ÷ total experiment}
\end{tabular}
\end{table*}

Our experimental study produces a large-scale dataset that systematically characterizes the resilience of Kubernetes-based cloud-edge systems. As summarised in Table~\ref{tab:cluster-arch-summary}, this dataset encompasses nearly 12,000 fault-injection experiments, totaling over 57 million request-level records and approximately 30~GB of structured time-series logs. Each experiment covers a unique configuration across multiple operational dimensions, including cluster size, deployment mode, application architecture (monolithic vs.\ microservices), fault type and intensity, and workload pattern, ensuring a broad and representative parameter space. Each experiment yields structured, per-request records, capturing request timestamps, application response times, average response times (both overall and for successful requests), failure rates, and detailed failure classification. This rich data collection enables fine-grained analysis and robust cross-factor evaluation of system behaviour under diverse real-world scenarios and injected failures.

The summary statistics in Table~\ref{tab:cluster-arch-summary} highlight several key findings. First, scaling cluster resources yields significant improvements in both performance and reliability: 8-node clusters consistently achieve lower mean and tail response times and reduced experiment failure rates compared to their 4-node counterparts. Second, architectural choice plays a critical role: microservices applications exhibit substantially higher tail latencies and failure rates than monolithic designs, even when controlling for cluster scale and injected fault characteristics. This performance gap underscores the heightened sensitivity of microservices to cascading failures and delay amplification under stress. These insights confirm that infrastructure capacity and application decomposition strategy shape resilience in cloud-edge environments. 

Our dataset establishes a rigorous foundation for benchmarking, comparative studies, and future advances in resilient distributed systems by providing fine-grained empirical coverage across environments, architectures, and failure types. The dataset is hosted in a private repository to support controlled access during the pre-publication phase. All materials will be made publicly available upon publication of the associated paper to facilitate reproducibility and accelerate research progress in the community. Interested readers may contact the author to request early access for academic purposes.

\subsection{Results and Analysis}

This section presents our comprehensive analysis of Kubernetes resilience in cloud-edge environments derived from systematic fault injection experiments. Our evaluation methodology employs dual-metric analysis, combining absolute performance measurements with normalized resilience indicators to reveal fundamental architectural trade-offs in distributed system design. We demonstrate that deployment decisions involve complex performance-resilience trade-offs rather than simple speed comparisons, with implications for mission-critical cloud-edge application design.

\subsubsection{Experimental Methodology and Normalization Framework}

To enable rigorous comparison across heterogeneous deployment environments with divergent baseline characteristics, we implement z-score normalization for all resilience metrics~\cite{Mutiny,kim2025impact}:
\begin{equation}
z = \frac{x - \mu}{\sigma} 
\end{equation}
where $x$ represents observed response time under fault conditions, $\mu$ denotes baseline mean response time, and $\sigma$ represents baseline standard deviation. Baselines are established using 25\% fault intensity measurements, representing the minimum perturbation level that triggers measurable system response while maintaining statistical validity across all experimental scenarios. This normalization approach centers each deployment configuration at $z = 0$ for baseline conditions and quantifies performance degradation in standard deviation units. Values of $z = 2$ indicate response times exceeding baseline by two standard deviations, signifying substantial performance degradation relative to environment-specific normal operation. Our dual-metric methodology, combining absolute response time analysis with normalized z-score evaluation—enables differentiation between environments that exhibit inherent performance characteristics versus those demonstrating volatility under stress conditions.

\begin{figure*}[b]
\centering
\begin{subfigure}[b]{0.53\linewidth}
    \centering
    \includegraphics[width=\linewidth]{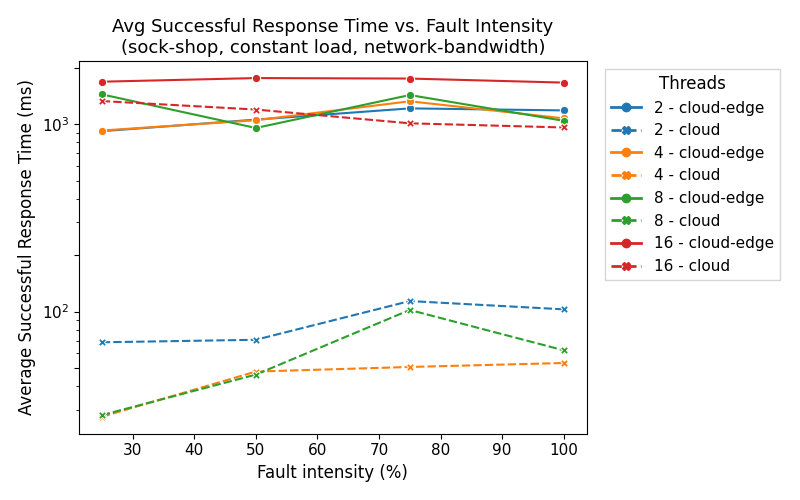}
    \caption{Network Bandwidth - Average Response Time}
    \label{fig:bandwidth-rt}
\end{subfigure}
\hfill
\begin{subfigure}[b]{0.46\linewidth}
    \centering
    \includegraphics[width=\linewidth]{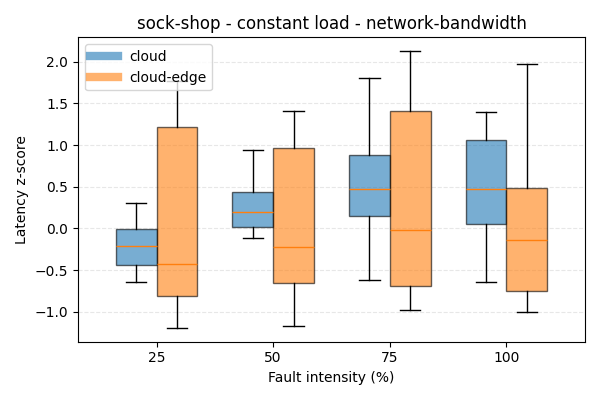}
    \caption{Network Bandwidth - Z-score Distribution}
    \label{fig:bandwidth-zscore}
\end{subfigure}
\caption{Resilience Metrics under Network Bandwidth Limitation}
\label{fig:bandwidth-combined}
\end{figure*}

\subsubsection{Network Fault Impact Analysis}

To systematically evaluate resilience characteristics across varied network disruption scenarios, we conduct comprehensive fault injection experiments spanning four distinct categories of network faults. Our systematic evaluation reveals distinct resilience patterns across four network fault categories, as summarized in Table~\ref{tab:fault-characteristics}. Each fault type demonstrates characteristic performance-resilience trade-offs that inform deployment decisions for cloud-edge environments.

\begin{table*}[t]
\centering
\caption{Network Fault Resilience Characteristics Summary}
\label{tab:fault-characteristics}
\small
\setlength{\tabcolsep}{6pt}
\renewcommand{\arraystretch}{1.15}
\begin{tabular}{lcll}
\toprule
\textbf{Fault Type} & \textbf{Absolute Speed$^\dagger$} & \textbf{Relative Resilience (Edge)$^\ddagger$} & \textbf{Preferred Deployment} \\
\midrule
Bandwidth Limitation & Slower & More volatile response (high $z$-variance) & Cloud \\
Network Loss          & Slower & More volatile response (high $z$-variance) & Cloud \\
Network Delay        & Slower & More stable response (low $z$-variance)    & Edge \\
Network Partition    & Slower & More stable response (low $z$-variance)    & Edge \\
\bottomrule
\multicolumn{4}{l}{\scriptsize $^\dagger$Absolute speed = baseline response time trend under fault conditions.}\\
\multicolumn{4}{l}{\scriptsize $^\ddagger$Relative resilience = stability of edge deployment compared to cloud (based on z-score variance).}
\end{tabular}
\end{table*}

These experimental results demonstrate the effectiveness of our dual-metric evaluation methodology in differentiating absolute performance characteristics from relative stability properties. The z-score normalization framework enables rigorous cross-environment comparison despite divergent baseline performance, revealing that edge deployments consistently exhibit slower absolute performance across all fault scenarios, but demonstrate contrasting resilience characteristics depending on fault type. Specifically, bandwidth limitation and network loss scenarios favor cloud deployments due to edge instability, while network delay and partition scenarios favor edge deployments due to superior relative stability.

\subsubsection{Bandwidth Limitation Effects}
Network bandwidth throttling experiments reveal fundamental differences in cloud versus edge resilience characteristics, as illustrated in Figure~\ref{fig:bandwidth-combined}. We systematically reduce available bandwidth from 25\% to 100\% fault intensity to evaluate how throughput constraints affect deployment resilience.

Absolute performance analysis in Figure~\ref{fig:bandwidth-rt} shows that both environments experience performance degradation under bandwidth constraints, though with different baseline characteristics. Under 16 concurrent users, bandwidth reduction from 25\% to 75\% increases cloud response times from approximately 1.7 seconds to 2.5 seconds, while edge response times surge from 2.3 seconds to 2.6 seconds. While edge deployments consistently exhibit higher absolute response times, both environments demonstrate similar degradation slopes under increasing bandwidth stress. However, the critical distinction emerges from z-score distribution analysis in Figure~\ref{fig:bandwidth-zscore}, which reveals that absolute performance degradation tells only part of the resilience story. Edge deployments (orange distributions) exhibit significantly larger interquartile ranges and extended whiskers than cloud deployments (blue distributions), with peak z-scores reaching $1.5\sigma$ versus cloud's $0.5\sigma$. This disparity indicates that while both environments suffer performance penalties, bandwidth limitations cause edge response times to deviate $1$–$2$ standard deviations from their baseline. In contrast, cloud environments maintain relative stability within $\pm1\sigma$ bounds despite experiencing similar absolute degradation.

This stability differential stems from fundamental architectural differences between deployment environments. Edge nodes' constrained network interface capabilities and limited buffer resources amplify bandwidth throttling effects, creating response time volatility that extends well beyond baseline performance characteristics. In contrast, cloud data centers leverage higher aggregate bandwidth and optimized network stacks, providing inherent resistance to throughput degradation. Our z-score normalization methodology proves essential for revealing this 2- 3x difference in response time variance under bandwidth constraints, a critical stability insight that conventional mean response time analysis would completely obscure.

\begin{figure*}[b]
\centering
\begin{subfigure}[b]{0.53\linewidth}
    \centering
    \includegraphics[width=\linewidth]{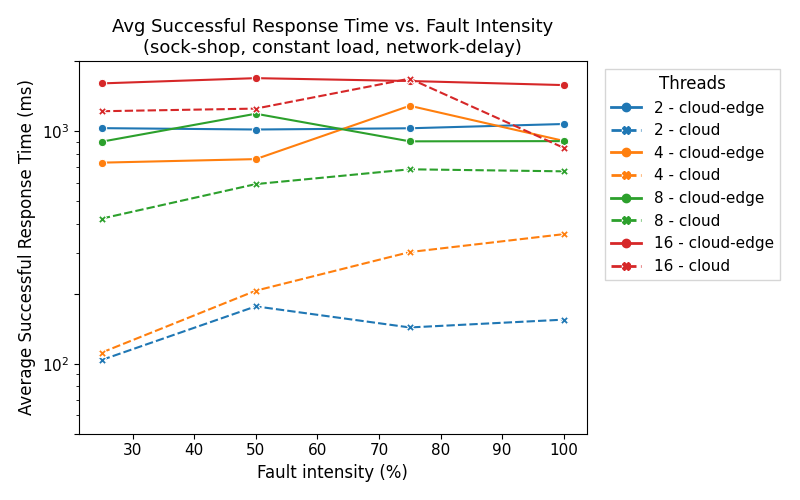}
    \caption{Network Delay - Average Response Time}
    \label{fig:delay-rt}
\end{subfigure}
\hfill
\begin{subfigure}[b]{0.46\linewidth}
    \centering
    \includegraphics[width=\linewidth]{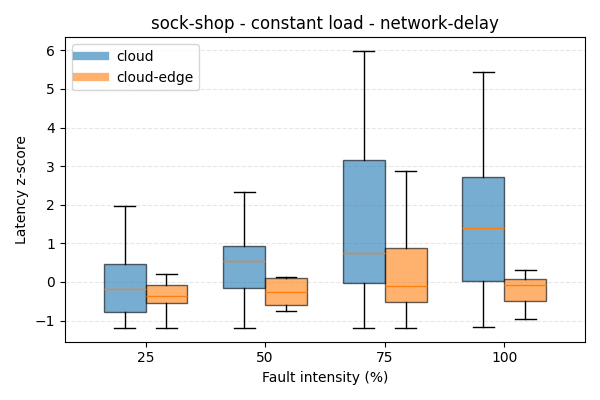}
    \caption{Network Delay - Z-score Distribution}
    \label{fig:delay-zscore}
\end{subfigure}


\begin{subfigure}[t]{0.53\linewidth}
    \centering
    \includegraphics[width=\linewidth]{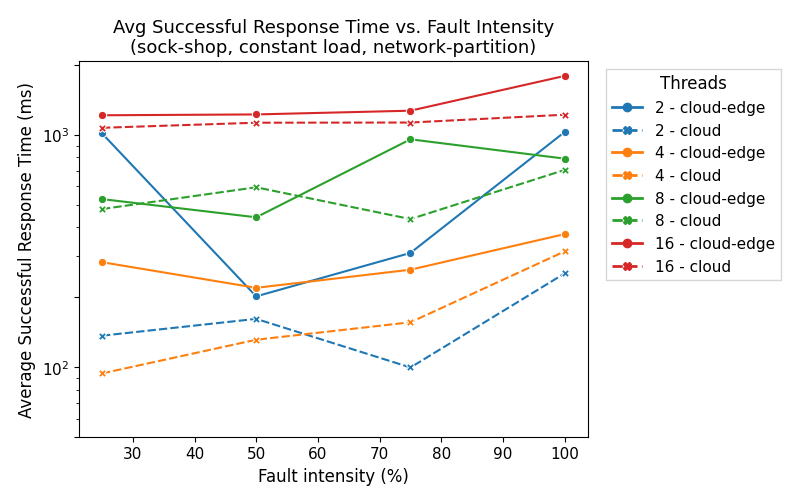}
    \caption{Network Partition - Average Response Time}
    \label{fig:partition-rt}
\end{subfigure}
\hfill
\begin{subfigure}[t]{0.46\linewidth}
    \centering
    \includegraphics[width=\linewidth]{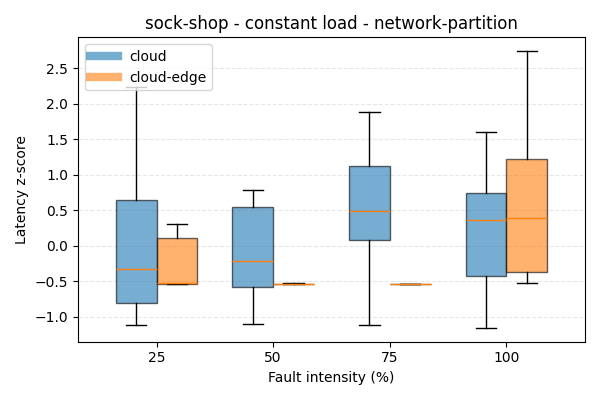}
    \caption{Network Partition - Z-score Distribution}
    \label{fig:partition-zscore}
\end{subfigure}

\caption{Resilience Metrics under Network Delay and Partition Faults}
\label{fig:delay-partition-combined}
\end{figure*}

\subsubsection{Network Delay Injection Analysis}

Network delay injection experiments reveal a fundamental resilience inversion between cloud and edge deployments, as demonstrated in Figure~\ref{fig:delay-partition-combined}. To evaluate latency sensitivity across deployment architectures, we systematically inject artificial delays ranging from 25\% to 100\% fault intensity into network communication paths.

While edge environments maintain consistently elevated absolute response times due to inherent latency penalties (approximately 200ms baseline), the relative resilience analysis tells a remarkably different story. As shown in Figure~\ref{fig:delay-rt}, both environments experience performance degradation under delay injection, but Figure~\ref{fig:delay-zscore} reveals dramatic differences in stability characteristics. Cloud z-score distributions expand substantially under delay injection, with median values reaching 3.0$\sigma$ and whiskers extending to 6$\sigma$ at 75-100\% fault intensity. Contrastingly, edge deployments remain tightly clustered within $\pm$1$\sigma$ ranges across all delay intensities.

This counterintuitive resilience inversion occurs due to fundamental architectural differences in communication patterns. Edge service invocations traverse fewer network hops and avoid extended cross-datacenter communication paths that suffer compounded delay effects. However, Cloud microservice architectures rely on multi-hop RPC chains that amplify injected latency as delays cascade through service dependencies, driving response times to multiple standard deviations beyond baseline performance. Edge deployments benefit from shorter local network paths that provide inherent protection against delay-based performance degradation. This demonstrates how architectural proximity can compensate for absolute performance limitations through superior resilience characteristics. Our systematic fault injection methodology combined with z-score analysis was essential for revealing this counterintuitive resilience inversion pattern, which challenges conventional performance-first deployment strategies and provides quantitative evidence that proximity-based architectures can achieve superior stability under network delay conditions despite inherent performance trade-offs.

\subsubsection{Network Partition and Network Loss Resilience}

Network connectivity disruption experiments examine two distinct failure modes that affect cloud and edge deployments differently: network partitions that fragment cluster connectivity and network loss that creates intermittent communication failures.

Network partition experiments, illustrated in Figure~\ref{fig:partition-rt} and Figure~\ref{fig:partition-zscore}, demonstrate clear edge resilience advantages under connectivity fragmentation scenarios. Edge deployments maintain z-scores below 0.5$\sigma$ across all partition intensities, while cloud medians climb to 1.1$\sigma$ at 75\% intensity. This stability differential occurs because edge services with local data caches and minimal cross-node dependencies remain largely unaffected by partial network segmentation. In contrast, cloud microservice chains must implement complex rerouting or experience stalling during partition events, resulting in measurable relative performance degradation as service dependencies become unreachable.

\begin{figure*}[t]
\centering
\begin{subfigure}[b]{0.53\textwidth}
    \centering
    \includegraphics[width=\textwidth]{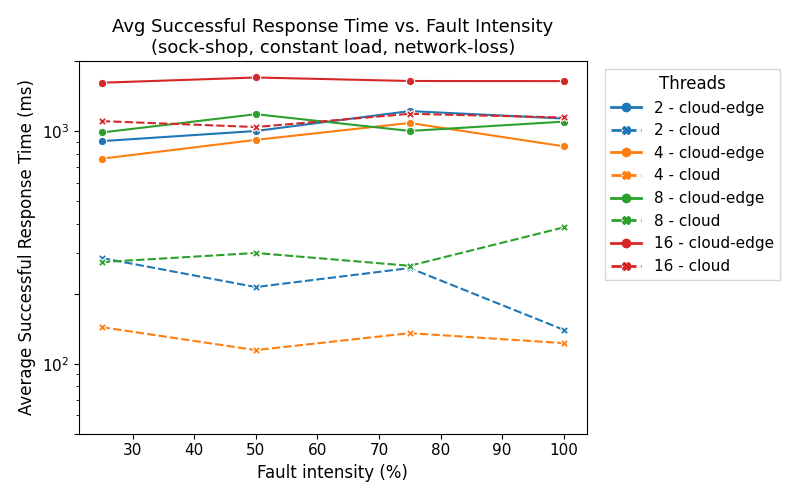}
    \caption{Network Loss - Avg Response Time}
    \label{fig:loss-rt}
\end{subfigure}
\hfill
\begin{subfigure}[b]{0.46\textwidth}
    \centering
    \includegraphics[width=\textwidth]{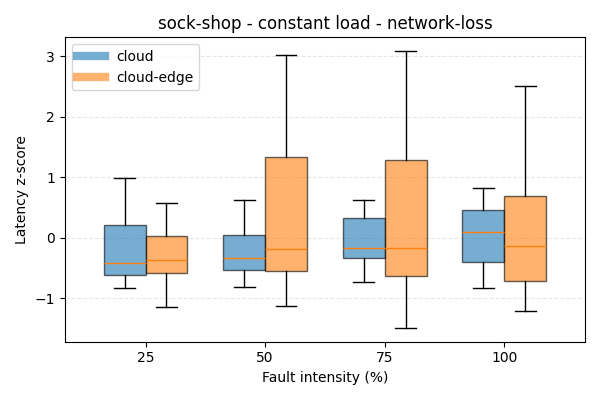}
    \caption{Network Loss - Z-score Distribution}
    \label{fig:loss-zscore}
\end{subfigure}

\caption{Resilience Metrics under Network Loss Faults}
\label{fig:partition-loss-combined}
\end{figure*}

However, network loss simulation reveals a contrasting pattern where cloud deployments demonstrate superior resilience. As shown in Figure~\ref{fig:partition-loss-combined}, network loss creates complex response patterns characterized by non-monotonic behavior, with average response times peaking at moderate loss levels (75\%) before declining at maximum intensity (100\%). This counterintuitive pattern reflects survivorship bias, where complete network loss causes most requests to fail or timeout, leaving only exceptional successful requests that skew performance averages downward. The z-score analysis reveals the underlying resilience characteristics: at moderate loss levels (50-75\%), edge median z-scores climb above 1$\sigma$ with whiskers reaching 3$\sigma$, while cloud environments maintain 0-0.6$\sigma$ ranges, indicating superior cloud resilience under partial network loss conditions where retry mechanisms and redundant communication paths provide stability advantages.

\subsection{Key Findings and Deployment Guidelines}

Synthesizing findings across our comprehensive fault injection experiments, we establish a systematic framework for evidence-based deployment decisions in cloud-edge environments. Our analysis of four distinct network fault categories reveals two fundamental performance-resilience patterns that define optimal deployment strategies for different operational scenarios.

The first pattern emerges under throughput-constrained scenarios involving bandwidth limitations and network loss conditions. Here, edge deployments demonstrate both slower baseline performance and higher volatility (large z-variance), while cloud deployments provide faster baseline performance with superior stability (smaller z-variance). These conditions consistently favor cloud deployment strategies for applications requiring consistent throughput and stable performance under adverse network conditions, where cloud infrastructure's optimized network stacks and redundant communication paths provide measurable resilience advantages.

Conversely, the second pattern emerges under latency-sensitive scenarios involving network delay and partition conditions. In these scenarios, edge deployments exhibit slower baseline performance but demonstrate superior relative stability (tight z-distributions), while cloud deployments provide faster baseline performance with higher volatility (large z-variance). These conditions favor edge deployment strategies for applications prioritizing predictable response characteristics over peak performance, where architectural proximity and reduced communication complexity provide inherent protection against latency amplification effects. This empirical analysis establishes that cloud-edge deployment decisions involve fundamental performance-resilience trade-offs rather than simple speed optimization. Edge deployments excel in providing stable, predictable service delivery under adverse network conditions, particularly when latency variations and intermittent connectivity represent primary operational challenges. Cloud deployments maximize absolute performance capabilities when network conditions remain favorable and applications can leverage high-bandwidth inter-component communication patterns without experiencing cascading delay effects.

Our empirical findings support evidence-based deployment decisions by systematically considering dominant network hazard types, application priority requirements (stability versus peak performance), and operational tolerance characteristics. Since no single cloud-edge architecture optimizes all failure modes simultaneously, resilience strategies must align with environment-specific network threat models and application-specific performance requirements. This comprehensive resilience framework provides the first quantitative foundation for context-aware deployment decisions by systematically identifying performance-resilience trade-off patterns that enable practitioners to move beyond intuition-based architectural selection toward evidence-driven deployment strategies.

\section{Conclusion}
\label{Conclusion}
This research develops a novel orchestration framework that automates fault injection, workload generation, and result collection across distributed cloud-edge Kubernetes environments. The framework enables systematic resilience experiments with one-click deployment capabilities across heterogeneous infrastructures. Using this framework, we conducted 11,965 fault injection experiments spanning pod-level, node-level, and network-level failures across both cloud and cloud-edge deployments. Our findings reveal two distinct fault response patterns: throughput-constrained scenarios favor cloud deployments with 47\% better resilience, while latency-sensitive scenarios favor edge deployments with 80\% superior response stability. Our results demonstrate that deployment decisions involve fundamental performance-resilience trade-offs, with optimal strategies depending on the dominant fault types in the target environment. 

Several limitations constrain our findings' generalizability, including controlled virtualized environments that may not capture full real-world complexity and simulated edge conditions. Future research directions include integrating additional fault injection and load generation platforms to enhance framework extensibility, validating findings in production environments with real edge deployments, developing more comprehensive fault coverage across different system layers, and improving the framework's modularity to support diverse experimental scenarios. This work establishes the foundation for evidence-based cloud-edge deployment strategies in mission-critical infrastructure.

\section{Ethics and Data Privacy}
\label{Ethics and Data Privacy}
Ethical approval and data privacy considerations are not required for this research, as it does not involve human participants, the collection of personal data, or any privacy-sensitive information. All fault injection experiments were performed on isolated research infrastructure using synthetic workloads and publicly available benchmark applications (Sock Shop, Image Detection), ensuring no impact on production environments or real user data. The resulting dataset contains only anonymized system-level performance metrics, such as response times, error rates, and resource utilization, with no personally identifiable information. The planned open-source release complies with best practices for reproducibility in computational research, providing only aggregated metrics suitable for scientific validation.

\subsection*{Author Contributions}
The author independently conducted all aspects of this study, including system design, Kubernetes setup, implementation, data analysis, and writing, using only the cited open-source tools.
\section*{Acknowledgment}
This research was supported by the NeCTAR Research Cloud, funded through the National Collaborative Research Infrastructure Strategy (NCRIS), which provided the computational infrastructure essential to this study.

\bibliographystyle{IEEEtran}
\bibliography{references}

\vfill
\clearpage
\onecolumn
\phantomsection
\addcontentsline{toc}{section}{Appendix: Code and Data Access}
\section*{Appendix: Code and Data Access}

\subsection*{A. Framework Code}
The complete implementation of the resilience testing framework is publicly available on GitHub for academic use:
\begin{flushleft}
\textbf{Repository URL:} \href{https://github.com/dylanC777/cloud-edge-k8s-resilience}{\texttt{https://github.com/dylanC777/cloud-edge-k8s-resilience}}
\end{flushleft}

\vspace{1em}

\subsection*{B. Dataset Availability}
Due to its large size, the dataset is not included in this thesis. Interested readers may contact the author to request access:
\begin{flushleft}
\textbf{Email:} \texttt{zche0292@student.monash.edu}
\end{flushleft}

\end{document}